\documentclass[12pt,a4paper]{article}

\begin{document}

\title{Precise study of three-nucleon systems within a representation without
explicit use of the isospin formalism}
\author{I.V.Simenog$^{a,b}$ , I.S.Dotsenko$^{b}$ , B.E.Grinyuk$^{a}$ \\
\\
$^{a}$\textit{Bogolyubov Institute for Theoretical Physics,} \\
\textit{Nat. Acad. of Sci. of Ukraine,} \textit{Kyiv 03143, Ukraine;} \\
(e-mail: isimenog@bitp.kiev.ua; bgrinyuk@bitp.kiev.ua)\\
$^{b}$ \textit{Taras Shevchenko National University,}\\
\textit{\ Kyiv 03127, Ukraine;} \\
(e-mail: dots@phys.univ.kiev.ua)}
\maketitle

\begin{abstract}
{\small A representation without explicit use of the isospin formalism is
developed for the precise study of few-nucleon systems, and the advantages
of the proposed approach are demonstrated. Using the example of
three-nucleon systems with central exchange NN-interaction potentials of the
general type, the complete equivalence is shown for the both approaches with
and without the isospin formalism. The new systems of equations contain a
less number of components as compared to the commonly used ones and are more
suitable for precise study of few-nucleon systems, in particular, within
variational approaches. Optimal variational schemes are developed with
Gaussian basis, and the binding energies, r.m.s. radii, density
distributions, formfactors, and pair correlation functions are calculated
with a high and controlled precision for }$^{3}${\small H and }$^{3}${\small %
He nuclei.}
\end{abstract}

\section{Introduction}

Noticeable progress in precise calculations of the few-nucleon binding
energies within variational approaches~\cite{1},~\cite{2},~\cite{3}
stimulates the further development of optimal variational schemes and their
possible refinement in order to achieve higher precision in calculations~
\cite{4},~\cite{5} of the main structural characteristics of the bound
states of few-nucleon nuclei. The progress in this field enables one to
obtain ''exact'' results for various structural characteristics of the
systems and to put forward the problem of precise fitting of the
nucleon-nucleon potentials universal for all the light nuclei, as well as to
carry out accurate variational calculations for heavier nuclei.

The achievement of a high accuracy in calculations of a few-particle quantum
system is restricted both by the number of particles and the number of the
wave function spatial components. At the first stage of studies of a simple
qualitative type, rather rough assumptions were used about nuclear
interaction potentials and the wave function spin structure (see~\cite{6},~
\cite{7},~\cite{8}), thus an account of all the wave function spatial
components was considered to be not necessary. But precise calculations with
controlled accuracy need to take into account all the factors contributing
to the result of a given accuracy. Such a statement of the problem is faced
with rather nontrivial difficulties even for comparatively simple three- and
four-nucleon systems. In the standard commonly used scheme with the use of
the isospin formalism, an additional symmetry of the wave function and the
mixture of states with different total isospins due to a consistent account
of the Coulomb interaction lead to the essential growth of the number of
spatial components of the wave function and thus to a serious complication
of calculations. As a result, to find the total wave function of the $^{3}$%
He nucleus in the framework of the isospin formalism, one has to solve a
system of six spatial equations, while in the case of $^{4}$He nucleus one
has already to deal with twelve equations.

Here, we present our approach proposed in~\cite{9} without explicit use of
the isospin representation and, using the example of three-nucleon systems,
show the complete equivalence of such a more convenient approach to the
traditional isospin formalism. It is shown for the wave functions in both
approaches to be connected by certain relations, while all the physical
observables calculated within both approaches are shown to coincide. The
obtained much more simple systems of equations for the spatial components of
the wave finctions of three nucleons in the bound state or scattering
process (in the doublet or quartet spin states) enabled us to carry out
precise calculations for three-nucleon nuclei and to analyze their structure.

\section{Equivalence of the representations with and without isospin}

In order to show the complete equivalence of the representation without
explicit use of isospin (protons are not identical to neutrons) to the
traditional isospin formalism, we recall the explicit form of equations for
the three-nucleon system ($2p,n$) within the isospin formalism with common
assumptions about the central exchange nuclear two-particle potential and
the Coulomb interaction between protons. In bound state ($^{3}$He nucleus),
the system ($2p,n$) has spin $S=1/2$ and the isospin projection $T_{3}=1/2$,
being in a mixed isospin state ($T=1/2$ and $T=3/2$). Thus, the total
antisymmetric wave function can be represented in terms of standard
spin-isospin functions and consists of six spatial components:

\begin{equation}\label{E1}
\Psi ^{a}=\psi ^{s}\xi ^{a}+(\psi ^{\prime }\xi ^{\prime \prime
}-\psi ^{\prime \prime }\xi ^{\prime })+\psi ^{a}\xi ^{s}+(\varphi
^{\prime }\zeta ^{\prime \prime }-\varphi ^{\prime \prime }\zeta
^{\prime })\chi _{3/2}\;.
\end{equation}

After projecting the Schr\"{o}dinger equation onto the basic spin-isospin
functions, one has the system of six equations for the spatial components of
the wave function:

\begin{eqnarray*}
&&\left[ \hat{K}-E+\frac{1}{2}\left( U_{31}^{s}+U_{13}^{s}\right) +\frac{1}{3%
}U_{c}^{s}\right] \psi ^{s}-\frac{1}{2}\left[ \left(
U_{31}^{\prime }-U_{13}^{\prime }\right) \psi ^{\prime }+\left(
U_{31}^{\prime \prime }-U_{13}^{\prime \prime }\right) \psi
^{\prime \prime }\right] \\ &&+\frac{1}{3}\left( U_{c}^{\prime
}\psi ^{\prime }+U_{c}^{\prime \prime }\psi ^{\prime \prime
}\right) +\frac{1}{3}\left( U_{c}^{\prime }\varphi ^{\prime
}+U_{c}^{\prime \prime }\varphi ^{\prime \prime }\right)\;=0,
\end{eqnarray*}

\begin{eqnarray*}
&&\left[ \hat{K}-E+\frac{1}{4}\left(
U_{33}^{s}+U_{31}^{s}+U_{13}^{s}+U_{11}^{s}\right) +\frac{1}{3}U_{c}^{s}%
\right] \psi ^{\prime } \\
&&-\frac{1}{4}\left[ \left( U_{33}^{\prime }-U_{31}^{\prime }-U_{13}^{\prime
}+U_{11}^{\prime }\right) \psi ^{\prime \prime }+\left( U_{33}^{\prime
\prime }-U_{31}^{\prime \prime }-U_{13}^{\prime \prime }+U_{11}^{\prime
\prime }\right) \psi ^{\prime }\right] \\
&&-\frac{1}{2}\left( U_{31}^{\prime }-U_{13}^{\prime }-\frac{2}{3}%
U_{c}^{\prime }\right) \psi ^{s}+\frac{1}{2}\left( U_{33}^{\prime \prime
}-U_{11}^{\prime \prime }+\frac{2}{3}U_{c}^{\prime \prime }\right) \psi ^{a}+%
\frac{1}{3}\left( U_{c}^{\prime \prime }\varphi ^{\prime
}+U_{c}^{\prime }\varphi ^{\prime \prime }\right)\;=0,
\end{eqnarray*}

\begin{eqnarray*}
&&\left[ \hat{K}-E+\frac{1}{4}\left(
U_{33}^{s}+U_{31}^{s}+U_{13}^{s}+U_{11}^{s}\right) +\frac{1}{3}U_{c}^{s}%
\right] \psi ^{\prime \prime } \\
&&-\frac{1}{4}\left[ \left( U_{33}^{\prime }-U_{31}^{\prime }-U_{13}^{\prime
}+U_{11}^{\prime }\right) \psi ^{\prime }-\left( U_{33}^{\prime \prime
}-U_{31}^{\prime \prime }-U_{13}^{\prime \prime }+U_{11}^{\prime \prime
}\right) \psi ^{\prime \prime }\right] \\
&&-\frac{1}{2}\left( U_{31}^{\prime \prime }-U_{13}^{\prime \prime }-\frac{2%
}{3}U_{c}^{\prime \prime }\right) \psi ^{s}-\frac{1}{2}\left( U_{33}^{\prime
}-U_{11}^{\prime }+\frac{2}{3}U_{c}^{\prime }\right) \psi ^{a}+\frac{1}{3}%
\left( U_{c}^{\prime }\varphi ^{\prime }-U_{c}^{\prime \prime
}\varphi ^{\prime \prime }\right)\;=0,
\end{eqnarray*}

\begin{eqnarray*}
&&\left[ \hat{K}-E+\frac{1}{2}\left( U_{33}^{s}+U_{11}^{s}\right) +\frac{1}{3%
}U_{c}^{s}\right] \psi ^{a}+\frac{1}{2}\left[ \left(
U_{33}^{\prime \prime }-U_{11}^{\prime \prime }\right) \psi
^{\prime }-\left( U_{33}^{\prime }-U_{11}^{\prime }\right) \psi
^{\prime \prime }\right] \\ &&-\frac{1}{3}\left( U_{c}^{\prime
}\psi ^{\prime \prime }-U_{c}^{\prime \prime }\psi ^{\prime
}\right) +\frac{1}{3}\left( U_{c}^{\prime }\varphi ^{\prime \prime
}-U_{c}^{\prime \prime }\varphi ^{\prime }\right) \;=0,
\end{eqnarray*}

\begin{eqnarray*}
&&\left[ \hat{K}-E+\frac{1}{2}\left( U_{13}^{s}+U_{33}^{s}\right) +\frac{1}{3%
}U_{c}^{s}\right] \varphi ^{\prime }+\frac{1}{2}\left[ \left(
U_{13}^{\prime }-U_{33}^{\prime }\right) \varphi ^{\prime \prime
}+\left( U_{13}^{\prime \prime }-U_{33}^{\prime \prime }\right)
\varphi ^{\prime }\right] \\ &&+\frac{1}{3}\left( U_{c}^{\prime
}\psi ^{\prime \prime }+U_{c}^{\prime \prime }\psi ^{\prime
}+U_{c}^{\prime }\psi ^{s}-U_{c}^{\prime \prime }\psi ^{a}\right)
\;=0,
\end{eqnarray*}

\begin{eqnarray}
&&\left[ \hat{K}-E+\frac{1}{2}\left( U_{13}^{s}+U_{33}^{s}\right) +\frac{1}{3%
}U_{c}^{s}\right] \varphi ^{\prime \prime }+\frac{1}{2}\left[
\left( U_{13}^{\prime }-U_{33}^{\prime }\right) \varphi ^{\prime
}-\left( U_{13}^{\prime \prime }-U_{33}^{\prime \prime }\right)
\varphi ^{\prime \prime }\right]  \\ &&+\frac{1}{3}\left(
U_{c}^{\prime }\psi ^{\prime }-U_{c}^{\prime \prime }\psi ^{\prime
\prime }+U_{c}^{\prime \prime }\psi ^{s}+U_{c}^{\prime }\psi
^{a}\right) \;=0,  \nonumber
\end{eqnarray}
\label{E2}where $\hat{K}$\ is the kinetic energy of three nucleons and

\begin{equation}
U^{s}=V(r_{12})+V(r_{13})+V(r_{23}),\;\;\; U^{\prime }=\frac{\sqrt{3}%
}{2}(-V(r_{13})+V(r_{23})),\;\;\; U^{\prime \prime }=-V(r_{12})+\frac{%
1}{2}(V(r_{13})+V(r_{23}))  \label{E3}
\end{equation}
are the combinations of spatial components of the potentials (both nuclear
and Coulomb ones) with a certain symmetry with respect to permutations of
nucleons. The last two equations in (\ref{E2}) are present due to the mixing
of different isospin states by Coulomb interaction. If we set $U_{c}=0$ in (%
\ref{E2}), we get the standard system of four equations~\cite{8} for $^{3}$%
He nucleus in both the spin ($S=1/2$) and isospin ($T=1/2$) doublet states,
as well as a separate system of two equations for the three-nucleon system
with $S=1/2$ and $T=3/2$.

The system of six spatial equations (\ref{E2}) is the basic one for studying
the $^{3}$He nucleus (as well as the scattering and reactions in the system (%
$2p,n$) ) within the isospin formalism. In practice, however, precise
calculations with the use of (\ref{E2}) have some difficulties because of a
considerable number of equations. We reformulate the problem in an
equivalent form using the fact that the reason for an increase in the number
of equations in the few-nucleon problem is the additional antisymmetrization
of wave functions (the generalized Pauli principle for nucleons) connected
with the introduction of isospin for treating a proton and a neutron as an
isotopic doublet. It appears that not all the spatial components of the wave
function in (\ref{E1}) are independent, and Eqs. (\ref{E2}) contain an
implicit relation between them. The analysis of system (\ref{E2}) shows that
it is suitable to introduce six new components $\phi _{1}^{\prime },$ $\phi
_{2}^{\prime },$ $\phi _{3}^{\prime },$ $\phi _{1}^{\prime \prime },$ $\phi
_{2}^{\prime \prime },$ $\phi _{3}^{\prime \prime }$ instead of \ $\psi
^{s},\psi ^{\prime },\psi ^{\prime \prime },\psi ^{a},\varphi ^{\prime
},\varphi ^{\prime \prime }$\bigskip :

\begin{equation}
\left(
\begin{array}{c}
\phi _{1}^{\prime } \\
\\
\phi _{2}^{\prime } \\
\\
\phi _{3}^{\prime } \\
\\
\phi _{1}^{\prime \prime } \\
\\
\phi _{2}^{\prime \prime } \\
\\
\phi _{3}^{\prime \prime }
\end{array}
\right) =\left(
\begin{array}{cccccc}
-\frac{1}{2\sqrt{3}} & -\frac{1}{2} & \frac{1}{2\sqrt{3}} & -\frac{1}{2} & 0
& -\frac{1}{\sqrt{3}} \\
&  &  &  &  &  \\
-\frac{1}{2\sqrt{3}} & \frac{1}{2} & \frac{1}{2\sqrt{3}} & \frac{1}{2} & 0 &
-\frac{1}{\sqrt{3}} \\
&  &  &  &  &  \\
\frac{1}{\sqrt{3}} & 0 & -\frac{1}{\sqrt{3}} & 0 & 0 & -\frac{1}{\sqrt{3}}
\\
&  &  &  &  &  \\
\frac{1}{2} & \frac{1}{2\sqrt{3}} & \frac{1}{2} & -\frac{1}{2\sqrt{3}} &
\frac{1}{\sqrt{3}} & 0 \\
&  &  &  &  &  \\
-\frac{1}{2} & \frac{1}{2\sqrt{3}} & -\frac{1}{2} & -\frac{1}{2\sqrt{3}} &
\frac{1}{\sqrt{3}} & 0 \\
&  &  &  &  &  \\
0 & -\frac{1}{\sqrt{3}} & 0 & \frac{1}{\sqrt{3}} & \frac{1}{\sqrt{3}} & 0
\end{array}
\right) \cdot \left(
\begin{array}{c}
\psi ^{s} \\
\\
\psi ^{\prime } \\
\\
\psi ^{\prime \prime } \\
\\
\psi ^{a} \\
\\
\varphi ^{\prime } \\
\\
\varphi ^{\prime \prime }
\end{array}
\right)
\begin{array}{c}
\\
\\
\\
\\
\\
\\
\\
\\
\\
\\
.
\end{array}
\label{E4}
\end{equation}
Then, as a result of the unitary (orthogonal) transformation, we get a new
form of system (\ref{E2}) with only two explicitly independent equations
from the set of six ones. Respectively, only two spatial components of the
wave function are independent, the rest being expressible through these two
ones with the account of the permutations of coordinates:
\[
\phi _{1}^{\prime }\left( 123\right) =-\frac{1}{2}\phi _{3}^{\prime }\left(
321\right) +\frac{\sqrt{3}}{2}\phi _{3}^{\prime \prime }\left( 321\right) ,%
\;\;\;\; \phi _{1}^{\prime \prime }\left( 123\right) =\frac{\sqrt{3}%
}{2}\phi _{3}^{\prime }\left( 321\right) +\frac{1}{2}\phi _{3}^{\prime
\prime }\left( 321\right) ,
\]
\begin{equation}
\phi _{2}^{\prime }\left( 123\right) =-\frac{1}{2}\phi _{3}^{\prime }\left(
132\right) -\frac{\sqrt{3}}{2}\phi _{3}^{\prime \prime }\left( 132\right) ,%
\;\;\;\; \phi _{2}^{\prime \prime }\left( 123\right) =-\frac{\sqrt{3%
}}{2}\phi _{3}^{\prime }\left( 132\right) +\frac{1}{2}\phi _{3}^{\prime
\prime }\left( 132\right) .  \label{E5}
\end{equation}
We note an important fact that the unitary transformation (\ref{E2}) is of
universal form and does not depend on the specific form of interaction
potentials but only on the spin-isospin symmetry in (\ref{E1}), (\ref{E2}).
To find all the wave function components and to obtain the total
antisymmetric function (\ref{E1}) for the system ($2p,n$), it is sufficient
to solve the system of only two equations (instead of six ones) for two
independent components $\phi _{3}^{\prime }$ and $\phi _{3}^{\prime \prime }$
(further we introduce notations $\Phi _{1}\equiv \phi _{3}^{\prime }$ and $%
\Phi _{2}\equiv \phi _{3}^{\prime \prime }$).

On the other hand, the same functions $\Phi _{1}$ and $\Phi _{2}$ can be
found directly from the system of two equations formulated in an equivalent
representation which does not use the notion of isospin and treats a proton
and a neutron as different particles. In this case, the total antisymmetric
(in protons) function of the system ($2p,n$) with total spin $S=1/2$ looks
as follows:
\begin{equation}
\Psi \left( p_{1},p_{2},n_{3}\right) =\zeta ^{\prime }\Phi _{1}\left(
p_{1},p_{2},n_{3}\right) +\zeta ^{\prime \prime }\Phi _{2}\left(
p_{1},p_{2},n_{3}\right) ,  \label{E6}
\end{equation}
where $\zeta ^{\prime }$ and $\zeta ^{\prime \prime }$ are the two known
components of the spin function of three nucleons with total spin $S=1/2$.
The spatial function $\Phi _{1}\left( p_{1},p_{2},n_{3}\right) $ is
symmetric, while $\Phi _{2}\left( p_{1},p_{2},n_{3}\right) $ is
antisymmetric with respect to permutations of identical protons. Within the
representation free from the notion of isospin, the system of equations for $%
\Phi _{1}$ and $\Phi _{2}$ has the following form:
\begin{eqnarray*}
&&\left[ \hat{K}+\frac{e^{2}}{r_{12}}+V_{s\left( pp\right) }^{+}\left(
r_{12}\right) -E\right] \Phi _{1}\left( 123\right) +\frac{1}{8}%
\sum_{ij=13,23}\sum_{+,-}\left[ 3V_{t\left( np\right) }^{\pm }\left(
r_{ij}\right) +V_{s\left( np\right) }^{\pm }\left( r_{ij}\right) \right] %
\left[ 1\pm P\left( ij\right) \right] \Phi _{1}\left( 123\right) + \\
&&+\frac{\sqrt{3}}{8}\sum_{ij=13,23}\sum_{+,-}\left( -1\right) ^{i+j}\left[
V_{s\left( np\right) }^{\pm }\left( r_{ij}\right) -V_{t\left( np\right)
}^{\pm }\left( r_{ij}\right) \right] \left[ 1\pm P\left( ij\right) \right]
\Phi _{2}\left( 123\right) =0,
\end{eqnarray*}
\begin{eqnarray}
&&\left[ \hat{K}+\frac{e^{2}}{r_{12}}+V_{t\left( pp\right) }^{-}\left(
r_{12}\right) -E\right] \Phi _{2}\left( 123\right) +\frac{1}{8}%
\sum_{ij=13,23}\sum_{+,-}\left[ V_{t\left( np\right) }^{\pm }\left(
r_{ij}\right) +3V_{s\left( np\right) }^{\pm }\left( r_{ij}\right) \right] %
\left[ 1\pm P\left( ij\right) \right] \Phi _{2}\left( 123\right) +  \nonumber
\\
&&+\frac{\sqrt{3}}{8}\sum_{ij=13,23}\sum_{+,-}\left( -1\right)
^{i+j}\left[ V_{s\left( np\right) }^{\pm }\left( r_{ij}\right)
-V_{t\left( np\right) }^{\pm }\left( r_{ij}\right) \right] \left[
1\pm P\left( ij\right) \right] \Phi _{1}\left( 123\right) =0\;.
\label{E7}
\end{eqnarray}

In the system of equations (\ref{E7}) for $^{3}$He nucleus, we take into
account the dependence of charge-dependent nuclear potentials on spin and
parity in the orbital momentum of nucleon pairs. In the case of $^{3}$H
nucleus consisting of two neutrons and one proton with the total spin $S=1/2$%
, one has also Eqs. (\ref{E7}), but with exchanged indices $%
n\longleftrightarrow p$ and without the Coulomb potential. Thus, Eqs. (\ref
{E7}) completely determine the properties of three nucleons in the doublet
spin state ($S=1/2$) both in the bound states and scattering processes.

The proposed approach is also applicable to the system of three nucleons in
the quartet state ($S=3/2$). All the properties of the system are determined
in this case by only the triplet charge-dependent nuclear interaction
potentials, and it is necessary to solve only one equation to find the
spatial component of the wave function (the spin function of three nucleons
is one-component in this case, and the spatial function of the system ($2p,n$%
) is antisymmetric with respect to permutations of the two protons). The
equation looks as follows:
\begin{equation}
\left[ \hat{K}+\frac{e^{2}}{r_{12}}+V_{t\left( pp\right) }^{-}\left(
r_{12}\right) -E\right] \Phi \left( 123\right) +\frac{1}{2}%
\sum_{ij=13,23}\sum_{+,-}V_{t\left( np\right) }^{\pm }\left(
r_{ij}\right) \;\left[ 1\pm P\left( ij\right) \right] \Phi \left(
123\right) =0\;.  \label{E8}
\end{equation}
It should be noticed that the complete equivalence of the systems with
different numbers of equations in different representations does not seem to
be a miracle if one rewrites the nuclear interaction operator in an
equivalent form in terms of the Majorana exchange operator without explicit
introduction of isospin operators. It follows from the general form of the
central exchange interaction potentials presented in terms of the Majorana
exchange operators that the transition to the equivalent representation
without use of the isospin formalism is also possible for heavier nuclei and
is reasonable for simplifying numerical calculations without any
approximations.

\section{Results of calculations}

To study the properties of three-nucleon systems using the system of Eqs. (%
\ref{E7}), we develop an optimal variational scheme \cite{5} with the use of
a Gaussian basis. Precise calculations are carried out for the basic
characteristics of the bound states of $^{3}$H and $^{3}$He nuclei for
various nuclear interaction potentials. Table 1 contains the results of
calculation of the binding energies $B=-E_{0}$ and r.m.s. radii for the
Afnan-Tang (ATS3) and Minnesota potentials ($\hbar ^{2}/M=41.47$ $MeV\cdot $
$fm^{2}$ in calculations with equal masses of a proton and a neutron, the
Coulomb parameter being $e^{2}=1.44$ $MeV\cdot $ $fm$, and the parameters of
potentials are taken from~\cite{3}). The more realistic Afnan-Tang potential
with interaction only in even orbital states is denoted by AT-(S3)$^{+}$.
The results for energies and radii are given with the accuracy of one unit
in the last digit with high probability (the exact results for the binding
energies are slightly greater, but within not more than $1$ $KeV$).

High accuracy of the calculations and the convergence of the results with
spreading the variational basis are obtained with the use of a comparatively
small number of basis functions (about 60-100 ones for both the symmetric $%
\Phi _{1}$and antisymmetric $\Phi _{2}$\ functions). It is found that the
optimal way to achieve a given accuracy is to take into account about three
times more Gaussian components for $\Phi _{1}$ than those for $\Phi _{2}$.
Due to the approach based on system (\ref{E7}) without explicit use of
isospin, our numerical results are of higher precision than the known
Varga-Suzuki ones~ \cite{3}, although we used a less number of Gaussian
variational basis functions.

The greater the difference between the triplet and singlet interaction
potentials and the greater the short-range repulsion, the more essential is
the difference between the energies calculated exactly and those obtained in
the spinless approximation. The complete calculation of energies with
account of the difference in the neutron and proton masses ($M_{p}\neq M_{n}$%
) shows the binding energy of $^{3}$H to increase a little and that of $^{3}$%
He to slightly decrease, while r.m.s. radii are practically unchangeable.
For all used interaction potentials, the proton, neutron, and mass r.m.s.
radii indicate the distinct specific structures of $^{3}$H and $^{3}$He
nuclei. Namely, the proton radius in the $^{3}$H nucleus is essentially
smaller than the neutron one (something like the neutron ''halo''), while
the $^{3}$He nucleus reveals the inverse pattern (the proton ''halo''). The
explanation of this effect lies in the fact that the attraction of a pair of
nucleons in the singlet state is weaker than that in the triplet one.

The obtained wave functions, having a suitable Gaussian representation,
enabled us to calculate the main structure functions of the three-nucleon
nuclei for various nuclear potentials. Fig. 1 presents the proton density
distribution $\rho _{p}(r)$\ as well as the neutron one $\rho _{n}(r)$ for a
$^{3}$H nucleus in the case of the AT-(S3)$^{+}$ potential (the nucleons are
considered to be point-like particles). The peripheral neutron ''halo''
effect is seen distinctly due to the neutron density distribution being
somewhat more long-range than the proton one (see also the r.m.s. radii in
Table 1). At the same time, the proton density is essentially higher than
the neutron one at the center of the nucleus. We have almost the same
density distributions for the $^{3}$He nucleus (even from the quantitative
point of view), but with the exchanged protons and neutrons $%
p\leftrightarrow n$. The main reason for such a regularity lies in
the difference between the interactions in triplet and singlet
states, while the Coulomb repulsion plays a negligible role. Fig.
2 shows the charge formfactors for $^{3}$H and $^{3}$He nuclei
calculated with the same potential. Note that the formfactor
profile of $^{3}$He falls down already at small $q^{2}$ more
rapidly as compared to that of $^{3}$H, due to the greater radius
of the proton density distribution in $^{3}$He as compared with
that in $^{3}$H. And the formfactor of $^{3}$He changes its sign
at smaller $q^{2}$ (experimental value $q_{min,\; exp}^{2}=11.6$
$fm^{-2}$ for $^{3}$He) because of the essential role of
short-range correlations
between the protons. Fig. 3 presents the pair correlation functions of the $%
^{3}$H nucleus both for the pair of neutrons ( $g_{2,\;nn}(r)$)
and for the neutron-proton pair ( $g_{2,\;np}(r)$) in the case of
the AT-(S3)$^{+}$ potential. A noticeable difference between them
is connected with a difference in the interactions in triplet and
singlet states, and with the fact that, in the $^{3}$H nucleus,
the proton and a neutron interact mainly in the triplet state,
while the two neutrons do in the singlet one. For comparison, we
give also the deuteron wave function squared in the same figure
for the same interaction potential (i.e., the pair
correlation function of two nucleons in the triplet state; see the curve $%
(np)_{D}$ in Fig.3) . The high accuracy of the known approximation based on
two-particle correlation functions for the few-nucleon nuclei is confirmed
once again.

All the main conclusions are also valid for the Minnesota potential, in
particular, those about the above-mentioned likeness of pair correlation
functions of three- and two-particle systems. Moreover, for the both
considered potentials, the pair correlation functions of $^{3}$H are similar
to those of $^{3}$He (with the account of the substitution $n\leftrightarrow
p$ like for density distributions). The structure of the nuclei is
determined mainly by the triplet and singlet interaction potentials, while
the role of the Coulomb interaction is small and leads to a little and
almost proportional expansion of the $^{3}$He nucleus in comparison with $%
^{3}$H. But, in the case of the Minnesota potential having essentially less
short-range repulsion than the AT-(S3)$^{+}$ one, the dips in the
formfactors of $^{3}$H and $^{3}$He nuclei occur at greater $q^{2}$, and the
pair correlation functions have an essentially less decrease at short
distances.

\section{Conclusions}

Thus, using the example of three-nucleon systems with central interaction
potentials, we show the complete equivalence of the isospin formalism (with
the total wave function being antisymmetric in the space of spin, isospin,
and coordinate variables) and the proposed representation without explicit\
use of isospin (with the total wave function being antisymmetric in
identical nucleons in spin and spatial coordinate variables). The obtained
equations in the representation without use of isospin are unitary
equivalent to the standard approach using the isospin representation.

The proposed approach is much more suitable (due to the small number of
spatial equations that is determined by the dimension of a spin Young
scheme) and enables one to carry out a precise study (with given accuracy)
of three-nucleon systems in the bound states or scattering processes with
various central nuclear interaction potentials. Using certain optimization
schemes for the variational method with Gaussian basis, we carried out the
precise calculations of the bound states of three nucleons with several
nucleon-nucleon interaction potentials. The results are of higher precision
than those available in the literature. Density distributions, formfactors,
and pair correlation functions are calculated as well.

The advantages of the proposed approach can reveal themselves in precise
calculations of systems of four nucleons (when the total spin $S=0$, there
are two equations for spatial components), five nucleons (there are five
equations at $S=1/2$), and, perhaps, of six-nucleon systems (five equations
for spatial components in the both cases of $^{6}$He and $^{6}$Li nuclei).
The developed approach opens a real possibility to construct the realistic
potentials describing completely, with reasonable accuracy, the main
low-energy parameters of few-nucleon systems.

\bigskip \textbf{Figure captions}

Fig.1

Profiles of proton and neutron density distributions for $^{3}$H nucleus
(for the AT-(S3)$^{+}$ interaction potential). Curves (1), (2) depict $\rho
(r)$, while curves (3), (4) show $r^{2}\rho (r)$.

Fig.2

Charge formfactors for $^{3}$H and $^{3}$He nuclei (for the AT-(S3)$^{+}$
interaction potential).

Fig.3

Pair correlation functions $g_{2,np}(r)$ and $g_{2,nn}(r)$ for $^{3}$H
nucleus (for the AT-(S3)$^{+}$ interaction potential). For comparison, the
pair correlation function $g_{2,pp}(r)$ for $^{3}$He and the deuteron wave
function squared are shown.

\bigskip

\textbf{Table 1}

Calculated binding energies $B=-E_{0}$ ($MeV$) and r.m.s. radii ($fm$) for $%
^{3}$H and $^{3}$He nuclei. $R_{p}$, $R_{n}$, $R_{m}$ are the proton,
neutron, and mass density distribution radii, respectively. (Experimental
values: $E_{0}=-8.481$ $MeV$ and $R_{p}=1.57$ $fm$ for $^{3}$H; $%
E_{0}=-7.716 $ $MeV$ and $R_{p}=1.70$ $fm$ for $^{3}$He).

\bigskip

\bigskip
\begin{tabular}{cc|ccc|ccc}
\hline &  &  & $^{3}H$ &  &  & $^{3}He$ &  \\ \hline & \ \ \ \ \ \
\ \ \ \  & $AT-(S3)^+$ & $ATS3$ & $ Minnesota$ & $AT-(S3)^+$ &
$ATS3$ & $Minnesota$ \\ \hline
& $-E_{0}$   & $6.699$ & $6.699$ & $\;6.896$ & $\;5.998$ & $%
\;5.998$ & $6.165$ \\ spinless approx. &  $R_{p}$ & $1.738$ &
$1.738$ & $1.730$ & $1.772$ & $1.772$ & $1.766$ \\ &  $R_{n}$ &
$1.738$ & $1.738$ & $1.730$ & $1.756$ & $1.756$ & $1.750$ \\ &
$R_{m}$ & $1.738$ & $1.738$ & $1.730$ & $1.766$ & $1.766$ &
$1.760$ \\ \hline & $-E_{0}$   & $7.491$ & $7.616$ & $7.561$ &
$6.833$ & $6.963$ & $6.882$ \\
$\;\Phi _{2}=0$ &  $R_{p}$ & $1.614$ & $1.591$ & $1.613$ & $1.790$ & $%
1.784$ & $1.796$ \\
&  $R_{n}$ & $1.758$ & $1.753$ & $1.761$ & $1.632$ & $1.608$ & $1.632$ \\
&   $R_{m}$ & $1.712$ & $1.701$ & $1.713$ & $1.739$ & $1.727$ & $1.743$ \\
\hline
& $-E_{0}$   & $8.491$ & $8.765$ & $8.386$ & $7.833$ & $8.110$ & $7.711$ \\
\ total &  $R_{p}$ & $1.576$ & $1.546$ & $1.586$ & $1.780$ & $1.763$ & $1.798
$ \\
&  $R_{n}$ & $1.749$ & $1.733$ & $1.763$ & $1.593$ & $1.560$ & $1.605$ \\
&   $R_{m}$ & $1.693$ & $1.673$ & $1.706$ & $1.720$ & $1.698$ & $1.736$ \\
\hline
& $-E_{0}$   & $8.495$ & $8.769$ & $\;8.389$ & $7.826$ & $8.103$ & $%
7.706$ \\
total$\;(M_{p}\neq M_{n})$ &  $R_{p}$ & $1.576$ & $1.546$ & $1.586$ & $%
1.781$ & $1.764$ & $1.799$ \\
&  $R_{n}$ & $1.748$ & $1.733$ & $1.762$ & $1.593$ & $1.561$ & $1.605$ \\
&   $R_{m}$ & $1.693$ & $1.673$ & $1.705$ & $1.721$ & $1.699$ & $1.736$ \\
\hline
\ \cite{3} & $-E_{0}$ \  & $\ -$ & $8.753$ & $8.380$ & $\ -$ & $\ -$ & $\ -$
\\
&   $R_{m}$ & $\ -$ & $1.67$ & $1.698$ & $\ -$ & $\ -$ & $\ -$ \\ \hline
\end{tabular}

\bigskip

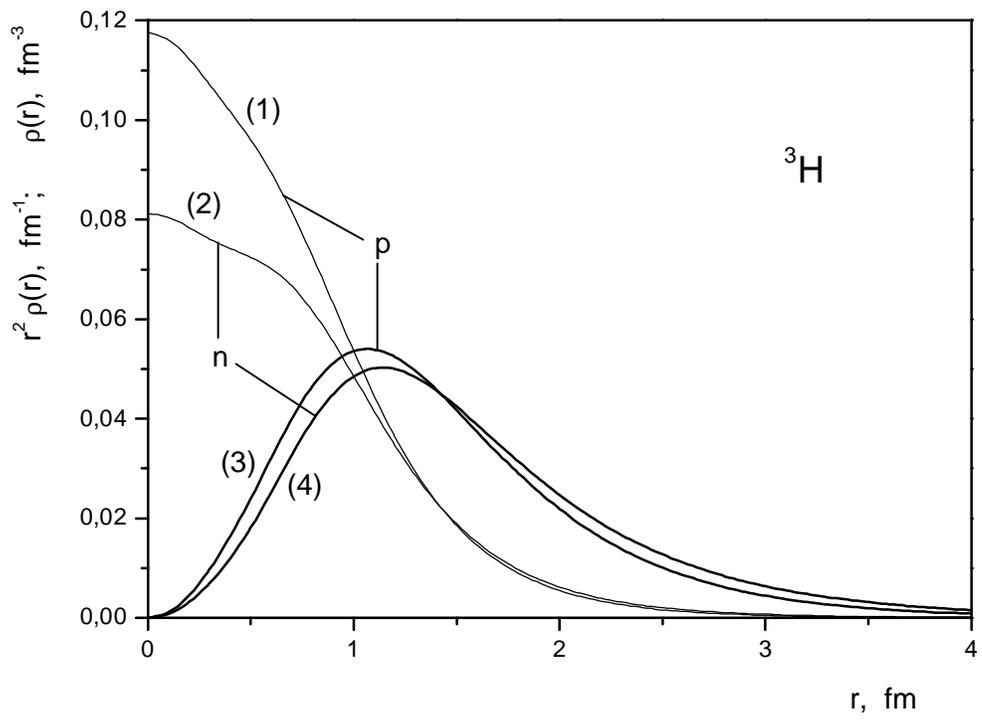
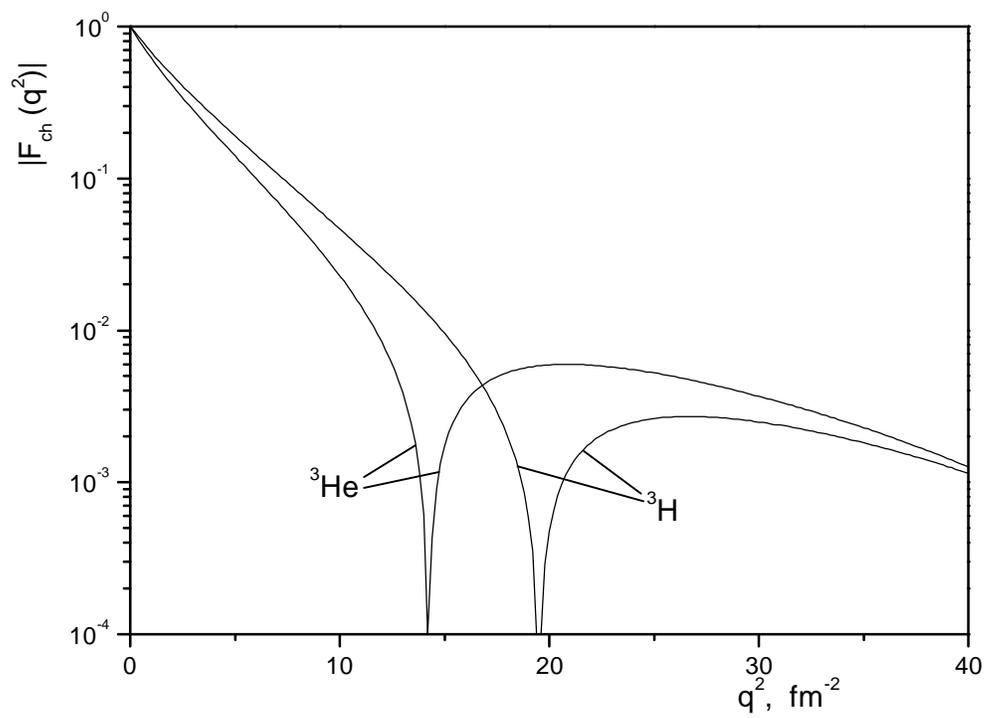
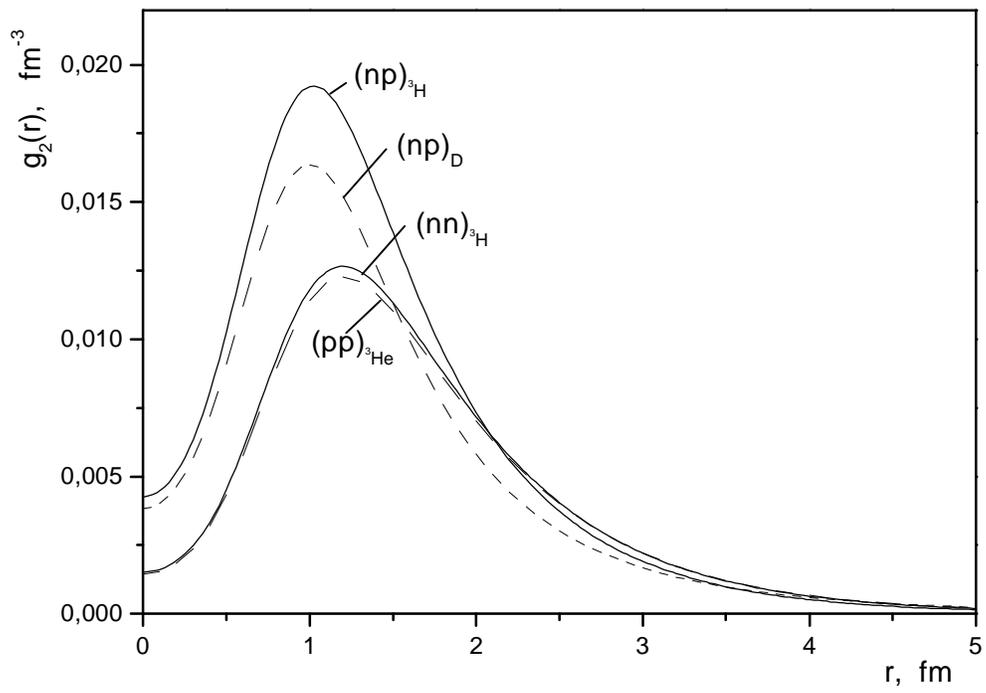
\end{document}